\documentclass[twocolumn,prl]{revtex4-2}

\usepackage{graphicx}
\usepackage{dcolumn}
\usepackage{bm}
\usepackage{multirow}
\usepackage{color}
\usepackage[normalem]{ulem}
\usepackage{amsmath}
\usepackage{gensymb}
\usepackage[colorlinks=true]{hyperref}

\usepackage{lineno}
\newcommand{\addMisha}[1]{\textcolor{blue}{#1}}

\begin{document}


\title{Interplay of vibrational, electronic, and magnetic states in CrSBr}
\author{Daria I. Markina$^{1}$}
\author{Priyanka Mondal$^{1}$}
\author{Lukas Krelle$^{1}$}
\author{Sai Shradha$^{1}$}
\author{Mikhail M. Glazov$^{2}$}
\author{Regine von Klitzing$^{1}$}
\author{Kseniia Mosina$^{3}$}
\author{Zdenek Sofer$^{3}$}
\author{Bernhard Urbaszek$^1$}
\email{bernhard.urbaszek@pkm.tu-darmstadt.de}

\affiliation{\small$^1$Institute for Condensed Matter Physics, TU Darmstadt, Hochschulstraße 6-8, D-64289 Darmstadt, Germany}
\affiliation{\small$^2$Ioffe Institute, 194021 St. Petersburg, Russia}
\affiliation{\small$^3$Department of Inorganic Chemistry, University of Chemistry and Technology Prague, Technicka 5, 166 28 Prague 6, Czech Republic}

\begin{abstract}

The van der Waals antiferromagnet CrSBr exhibits coupling of vibrational, electronic, and magnetic degrees of freedom, giving rise to distinctive quasi-particle interactions. We investigate these interactions across a wide temperature range using polarization-resolved Raman spectroscopy at various excitation energies, complemented by optical absorption and photoluminescence excitation (PLE) spectroscopy. Under 1.96 eV excitation, we observe pronounced changes in the A$_g^1$, A$_g^2$, and A$_g^3$ Raman modes near the N\'eel temperature, coinciding with modifications in the oscillator strength of excitonic transitions and clear resonances in PLE. The distinct temperature evolution of Raman tensor elements and polarization anisotropy for Raman modes indicates that they couple to different excitonic and electronic states. The suppression of the excitonic state's oscillation strength above the N\'eel temperature could be related to the magnetic phase transition, thereby connecting these excitonic states and Raman modes to a specific spin alignment.
These observations make CrSBr a versatile platform for probing quasi-particle interactions in low-dimensional magnets and provide insights for applications in quantum sensing and quantum communication.

\end{abstract}

\maketitle

\textbf{Introduction.---} 
A detailed investigation of quasi-particle interactions in quantum materials, involving charges, spins, excitons, and phonons, as well as the effects caused by these interactions, represents an important scientific challenge \cite{freeman1968exciton,datta2025magnon,liebich2025controlling,doi:10.1021/acs.nanolett.5c03150}. Understanding these phenomena is essential for progress in spintronics \cite{zhang20242d, gibertini2019magnetic}, quantum sensing \cite{radtke2019nanoscale, fang2024quantum, melendez2025quantum}, quantum communication, and quantum computing \cite{zhong2025integrating, younis2025magnetoresistance}.

Magnetic two-dimensional (2D) materials are particularly well-suited for exploring spin–phonon, electron-phonon, and exciton-phonon interactions \cite{park20252d}. For example, spin–phonon coupling has been studied in MPS$_3$ compounds (M = Ni, Fe, Co, Mn) \cite{rao2024mode} and in MnBi$_{2n}$Te$_{3n+1}$ (n = 1, 2, 3, 4) \cite{cho2022phonon}. It was established 
how specific phonon modes couple with magnetic subsystem both above and below the Néel temperature. For MPS$_3$ compounds, it was shown that the variation of the strength of the coupling is attributed to changes in metal–sulfur bond lengths. 

Recently, the novel van der Waals antiferromagnet with ferromagnetic intralayer spin alignment\cite{goser1990magnetic}, CrSBr, has revealed striking features of various quasi-particle interactions. Its unique quasi-1D crystal structure \cite{klein2023bulk} gives rise to strong anisotropy in electronic, magnetic, and optical properties \cite{wilson2021interlayer, tabataba2024doping}. Several studies have investigated this behavior. Temperature- and magnetic-field-dependent Raman spectroscopy \cite{pawbake2023raman, torres2023probing, wdowik2025magneto} reveals fingerprints of spin-phonon coupling, such as phonon modes intensity changes, appearance of new Raman modes due to the symmetry breaking caused by spin alignment, and higher-order scattering processes below the N{\'e}el temperature.
Exciton–phonon interactions are equally significant. Photoluminescence spectra at low temperatures \cite{lin2024strong} show discrete phonon sidebands resulting from exciton–phonon coupling. Ultrafast optical and electron diffraction experiments \cite{meineke2024ultrafast, ranhili2025ultrafast} further demonstrate how exciton dynamics and coherent phonon generation occur in a crystal with a specific magnetic order. However, many aspects of the mentioned interactions remain unclear.

A powerful tool to probe hidden material properties is polarization-resolved Raman spectroscopy \cite{pimenta2021polarized,ferrari2013raman}. As we showed in our earlier work at ambient conditions \cite{mondal2025raman}, Raman modes in CrSBr display strong sensitivity to excitation energy in the paramagnetic phase. 
Investigating Raman scattering and optical resonances in absorption as a function of temperature allows us here to explore further how magnetisation influences quasi-particle coupling in CrSBr.

\textbf{Experimental Results.---}
2D materials, including CrSBr, can be separated down to a monolayer by mechanical exfoliation\cite{liu2021mechanical} due to the weak van der Waals forces that hold the layers of material together in the out-of-plane dimension. The crystal exhibits a pronounced structural anisotropy: it is elongated along the $a$ crystallographic axis, known as the intermediate magnetic axis, while its shorter side corresponds to the $b$
axis, the magnetic easy axis (Figure \textcolor{red}{S1}). The out-of-plane direction corresponds to the $c$ axis, the hard magnetic axis \cite{wilson2021interlayer}. These crystallographic features define the electronic and optical properties of the material. The photoluminescence (PL) in CrSBr, from monolayer to bulk, is strongly polarized along the $b$ axis, reflecting the anisotropy of the band structure (Figure \textcolor{red}{S2}). 

Raman scattering also exhibits polarization dependence \cite{klein2023bulk}, but unlike PL, its polarization is sensitive to the excitation energy. In particular, altering the laser excitation energy enables switching of polarization direction between $a$ and $b$ axes of the Raman A$_g^2$ mode under ambient conditions \cite{mondal2025raman}, as can be seen in Fig.~\ref{fig:fig1}. A typical Raman spectrum of CrSBr under 1.96 eV excitation (Figure \textcolor{red}{S3}, top panel) displays three major peaks corresponding to optically active out-of-plane vibrational modes: A$_g^1$, A$_g^2$, and A$_g^3$ known from the literature\cite{klein2023bulk,torres2023probing,doi:10.1021/acs.nanolett.3c01920}. \\
\indent In our previous study \cite{mondal2025raman}, we showed that laser excitation energies of 1.96 eV and 2.33 eV lead to distinct behaviors of the A$_g^2$ mode. To provide a comparative study, we employed the same laser sources (see Methods Section) to excite a 5 nm thick sample region (blue circle in Figure \textcolor{red}{S1}), see data in Fig.~\ref{fig:fig1}. Data for other thicknesses is shown for comparison in Figs.~\ref{fig:fig2} and ~\ref{fig:fig3}. Polarization-resolved Raman measurements were performed in a closed-cycle cryostat with precise sample temperature control in a co-polarized configuration, Figure \textcolor{red}{S4}.

The electronic band structure also evolves with the sample temperature. In the magnetic semiconductor CrSBr, this evolution is closely intertwined with magnetic ordering. The system undergoes a transition from the paramagnetic (PM) phase to an intermediate ferromagnetic (iFM) phase at the Curie temperature (T$_C$ $\sim$165 K), where interlayer ordering is only partial, exhibiting both ferro- and antiferromagnetic interlayer couplings. Further temperature decrease leads to a transition to the antiferromagnetic (AFM) phase at the N\'eel temperature (T$_N$ $\sim$132 K) \cite{pei2024surface, pawbake2023raman} (Figure~\ref{fig:fig1}, right column). Therefore, Raman modes, electronic states, and magnetisation in CrSBr are coupled, which affects the optical response of the system. 
To investigate this interplay, we report detailed polarization-resolved Raman scattering measurements for different laser energies across a wide temperature range from 4 to 300 K, focusing on the interplay between phonons and spin alignment in CrSBr.

For our measurements under 2.33 eV excitation, the polarization of the A$_g^2$ and A$_g^3$ modes remains fairly constant down to $\sim$80 K: the A$_g^2$ mode is consistently polarized along the $a$ axis, while the A$_g^3$ mode aligns along the $b$ axis (Figure~\ref{fig:fig1}, center column). Below $\sim$80 K, however, the A$_g^2$ mode exhibits a notable change; its polar pattern acquires a slight four-fold symmetric component, indicating an additional signal component along the $b$ axis. In contrast, the A$_g^3$ mode shows no such variation across the entire temperature range of 4–250 K.

The response under 1.96 eV excitation is dramatically different. For both A$_g^2$ and A$_g^3$ modes, temperature strongly affects the polarization dependencies (Figure~\ref{fig:fig1}, left column). At room temperature, the A$_g^2$ mode is purely polarized along the $b$ axis, but as the temperature decreases, its polar pattern changes significantly, with the $a$ axis component growing until it becomes comparable to the $b$ axis intensity. The A$_g^3$ mode follows a similar evolution, but with the $a$-component eventually exceeding the $b$-component below the N\'eel temperature.
In both excitation cases (1.96 and 2.33 eV), the overall Raman intensity of all modes decreases with decreasing temperature.

The polar dependencies shown in Figure~\ref{fig:fig1} contain rich information that can be extracted by fitting the polar plots using the following expression \cite{pimenta2021polarized}:
\begin{equation}
    \label{fit}
    I_{||}^{A_g}({\theta})=(a{\cdot}\text{cos}^2{\theta}+b{\cdot}\text{cos}{\phi}_{ab}{\cdot}\text{sin}^2{\theta})^2+b^2{\cdot}\text{sin}^4{\theta}{\cdot}\text{sin}^2{\phi}_{ab}\addMisha{,}
\end{equation}
where parameters $a$ and $b$ correspond to the amplitudes along the $a$- and $b$-axes, respectively, and the phase $\phi_{ab}$ accounts for the imaginary part of the Raman tensor elements related to an interplay of virtual and real intermediate states (see Discussion Section for details).

As it is evident from the polar plots, for the A$_g^2$ mode, the $a$-component increases by about 1.5 times under 2.33~eV excitation and more than 6.5 times under 1.96~eV excitation (Figure \textcolor{red}{S5 a, b}). For the A$_g^3$ mode, the $a$-component remains negligibly small at 2.33 eV, but rises nearly fourfold under 1.96 eV excitation (Figure \textcolor{red}{S5 a, b}).
Unlike the relatively monotonous evolution of the $a$-component, the $b$-component shows a distinct kink near the N\'eel temperature (Figure \textcolor{red}{S5, c, d}). Importantly, this kink appears only for the A$_g^2$ mode under 2.33 eV excitation and only for the A$_g^3$ mode under 1.96 eV excitation. This suggests that, despite sharing the same symmetry, the A$_g^2$ and A$_g^3$ modes couple to different electronic states, which are selectively excited by different laser energies.

The absolute changes in all parameters are significantly more pronounced under 1.96 eV excitation. Given these stronger effects observed at 1.96 eV, our subsequent discussion focuses on this case.

To investigate the influence of band structure changes on (near) resonant Raman scattering, it is essential to trace the Raman modes in samples of different thicknesses. For this purpose, polarization-resolved Raman spectroscopy over a wide temperature range was performed at various regions of the same sample (Figure \textcolor{red}{S1}, markers denote different thickness regions). As a universal parameter illustrating the behavior of the Raman modes, we use the $b/a$ ratio.
Interestingly, each Raman mode exhibits a characteristic response feature near the N\'eel temperature across the broad thickness range (Figure~\ref{fig:fig2}). The A$_g^1$ mode shows a step-like change in the $b/a$ ratio at the N\'eel temperature, with the effect slightly more pronounced in thicker samples (Figure~\ref{fig:fig2}a). For the A$_g^2$ mode, no drastic change is visible at first glance (Figure~\ref{fig:fig2}b); however, its first derivative reveals a distinct peak at the magnetic phase transition (see Figure \textcolor{red}{S6}).

Our measurements show the most significant changes for the A$_g^3$ mode. Indeed, the $b/a$ ratio of the A$_g^3$ mode displays a clear kink at the N\'eel temperature for all tested thicknesses, though the kink becomes progressively less sharp as thickness increases in Fig.~\ref{fig:fig2}c. For the 5 nm sample, the kink in the $b/a$ ratio appears less pronounced, likely due to the extremely large $a$-component, which exceeds the $b$-component below the N\'eel temperature (see components plotted separately in Figure \textcolor{red}{S7}).  \\
\indent The Raman scattering results in Figs.~\ref{fig:fig1} and \ref{fig:fig2} show sensitivity of the vibrational modes to magnetic order. We need to further investigate if the coupling of the Raman modes to magnetisation is direct or mediated by electronic states excited by the laser in the Raman experiments. To achieve this target, we performed photoluminescence excitation spectroscopy (PLE) complemented by differential reflectance contrast (DR/R) measurements (Figure~\ref{fig:fig3}). The PLE signal was obtained as the integrated signal under the so-called bright (X$_A$) and dark (X$_D$) exciton peaks \cite{krelle2025magnetic} (Figure \textcolor{red}{S2}). The PLE and DR/R spectra show matching spectral features (peaks and dips) at similar energies. \\

\indent For the few-layer sample (5 nm thick) at 4 K, pronounced peaks appear at 1.77 eV and 1.82 eV under excitation polarized along the $b$ axis, while excitation along the $a$ axis produces a peak at 1.96 eV (Figure~\ref{fig:fig3}a,b). In the thicker 31 nm flake, the peak positions remain largely unchanged: excitation along the $a$ axis shows little variation, but along the $b$ axis, additional peaks emerge around $\sim$1.86 eV and $\sim$1.9 eV (Figure~\ref{fig:fig3}d,e), which might relate to the polaritonic effects \cite{li20242d,dirnberger2023magneto}.
No clear resonances are observed near 2.33 eV (see Figure \textcolor{red}{S8} for extended data). 
In contrast, a clear resonance consistently appears near 1.96 eV polarized along the $a$ axis for a range of thicknesses, perfectly matching the chosen excitation conditions for our Raman experiments. In thicker samples, the 1.96 eV excitation lies on the shoulder of an additional resonance. The resonant behaviour is additionally supported by the appearance of extra peaks between 370 and 780 cm$^{-1}$ under 1.96 eV excitation \cite{pawbake2023raman}, in contrast to the case of 2.33 eV excitation (Figure \textcolor{red}{S9}).  

The most pronounced PLE peak along the $b$ axis is the one at 1.77 eV, which corresponds exactly to the so-called B-exciton (X$_B$) \cite{smiertka2025unraveling, shi2025giant}, which couples more strongly than X$_A$ to magnetisation and phonons. Furthermore, the temperature dependence of DR/R under $b$ axis excitation shows that strong spectral features near 1.96 eV vanish above the N\'eel temperature, clearly indicating their coupling to magnetic order, see Fig.~\ref{fig:fig3}c, f., as has been discussed in DR/R for the resonances near the X$_A$ exciton \cite{shao2025magnetically}.

\textbf{Discussion.---} 
The intensity of the Raman scattering is directly determined by the Raman tensor $\hat R$ and polarization vectors of the incident, $\mathbf e_i$, and scattered, $\mathbf e_s$, fields as $I_s \propto |\mathbf e_i \hat R \mathbf e_s|^2$. The Raman tensor in turn depends on both the crystal symmetry and the more importantly, the electron-phonon coupling.
For the phonon modes of $A_g$-symmetry the non-zero components of the Raman tensor are $ae^{\mathrm i \phi_a}=R_{aa}$, $be^{\mathrm i \phi_b}=R_{bb}$, and $ce^{\mathrm i \phi_c}=R_{cc}$ where $a$, $b$, and $c$ are the principal axes of the system:
\begin{equation}
{R}(A_g)=
\begin{pmatrix}
a e^{i \phi_a} & 0 & 0 \\
0 & b e^{i \phi_b} & 0 \\
0 & 0 & c e^{i \phi_c}
\end{pmatrix}
\label{eq:my_matrix}
\end{equation}
Equation~\eqref{fit} follows from Eq.~\eqref{eq:my_matrix} with $\phi_{ab} = \phi_a - \phi_b$.
Given the large energy difference between excitation photons and scattered phonons, direct light–phonon coupling is unlikely. Instead, Raman scattering is mediated by electronic excitations. According to \cite{peter2010fundamentals}, the Raman tensor component for a phonon branch $\mu$ ($R_{ij}^{\mu}$) reflects the sequence of physical processes underlying Raman scattering: (i) virtual or real transition from the ground state of the crystal $|0\rangle$ to one of the excited electronic states $|n\rangle$, (ii) real process of the phonon emission with the transition between the excited states $|n\rangle$ and $|n'\rangle$, the state $|n'\rangle$ can be both real or virtual, and (iii) photon emission from the state $|n'\rangle$ returning the crystal in the ground state. The Raman tensor components are therefore expressed as:
\begin{widetext}
\begin{equation}
    \label{Raman}
    R_{ij}^{\mu}(E_L)=\sum_{n,n'}\frac{\langle 0|\mathbf e_s^*\cdot \hat{\mathbf{p}}|n'\rangle\langle n'|H_{el-ph}^{\mu}|n\rangle \langle n|\mathbf e_i\cdot \hat{\mathbf{p}}|0\rangle}{[E_L-E_{n}+i\Gamma_n][E_L-E_{ph}^{\mu}-E_{n'}+i\Gamma_{n'}]}\addMisha{,}
\end{equation}
\end{widetext}
where $E_{L}$ is the laser excitation energy, $E_{n}$, $E_{n'}$ are energies of intermediate electronic states  $|n\rangle$ and $|n'\rangle$, $\Gamma_n$, $\Gamma_{n'}$ are their damping, $\hat p$ is a momentum operator element and $H_{el-ph}^{\mu}$ stands for the Hamiltonian of the electron interaction with the phonon mode $\mu$ with the energy $E_{ph}^\mu$. The indices $i$ and $j$ denote the polarization states of the incident and scattered photons, respectively. It follows from the microscopic expression~\eqref{Raman} that variations of the Raman scattering intensity and polarization across the magnetic phase transitions can be related to the specifics of light-matter coupling of the intermediate states. Indeed, the magnetic order is coupled with electronic excitations affecting both the optical transition matrix elements $\langle n|\mathbf e_i\cdot \hat{\mathbf{p}}|0\rangle$, $\langle 0|\mathbf e_s^*\cdot \hat{\mathbf{p}}|n'\rangle$ and electron-phonon interaction $\langle n'|H_{el-ph}^{\mu}|n\rangle$. 

As shown in Fig.~\ref{fig:fig3}c,f, the linear optical response of CrSBr crystals in the energy range of 1.6 \ldots 2~eV  drastically changes in the vicinity of the N\'eel temperature, where the features related to the X$_B$ exciton practically vanish at $T>T_N$. It might be caused by both the significant enhancement of the linewidth broadening (increase of $\Gamma_{n,n'}$ in the denominator of Eq.~\eqref{Raman}) as well as the reduction of the oscillator strength (optical transition matrix elements in the numerator of Eq.~\eqref{Raman}) related to the magnetic disorder~\cite{shao2025magnetically}. Hence, the almost-resonant contributions to the Raman scattering at $E_L=1.96$~eV mediated, e.g., by X$_B$ exciton become suppressed at $T>T_N$, leading to changes of the magnitudes of $a$ and $b$ components and characteristic temperature-dependent evolution of the $b/a$ ratio, linking these excitonic states and hence Raman modes to a specific magnetic phase. Different behavior of $b/a$ amplitude ratio for $A_g^1$, $A_g^2$, and $A_g^3$ modes (Fig.~\ref{fig:fig2}) implies that different excitonic and electronic states underlie the light scattering by these modes. 

It is noteworthy that the phonon energies experience only a slight change across the phase transition (Figure \textcolor{red}{S10a}), suggesting that direct spin–phonon interaction may play a less significant role in the observed phenomena.


\indent \textbf{Conclusion.---} 
 Using temperature-dependent polarization-resolved Raman spectroscopy and measuring optical absorption, we demonstrate that, in addition to the previously reported excitation energy dependence \cite{mondal2025raman}, the Raman modes of CrSBr are also sensitive to the magnetic phase through the complex interplay of electronic, magnetic, and vibrational states. In particular, we observe clear polarisation changes for the three primary optically active Raman modes near the N\'eel temperature under 1.96 eV excitation and correlate these findings with optical absorption measurements. Temperature-dependent optical absorption combined with PLE spectroscopy reveals nearly resonant contributions to the Raman scattering, mediated by the high-energy transitions near the X$_B$ excitonic states, which become suppressed above the Néel temperature. The distinct temperature-dependent evolution of the $b/a$ ratio for different Raman modes underscores the role of specific excitonic and electronic states involved in light scattering processes for each mode. We propose that the observed connection between the magnetic phase and phonon modes (spin–phonon interaction) arises indirectly via electron–phonon (exciton–phonon) coupling. 
 Our findings highlight the intricate coupling among spin, lattice, excitonic, and electronic states in CrSBr.

\textbf{METHODS}\\
\indent\textbf{Sample Fabrication}\\
Bulk CrSBr crystals were fabricated through chemical vapor transport \cite{klein2022control}. The samples were prepared through mechanical exfoliation onto SiO$_2$/Si substrates with an 85 nm SiO$_2$ layer.\\
\indent\textbf{Atomic Force Microscopy}\\
To probe the topology and thickness of the CrSBr flakes atomic force microscopy measurements were performed at room temperature on a Cypher AFM (Asylum Research/Oxford Instruments, Wiesbaden, Germany). Height images of CrSBr flakes were obtained in AC tapping mode using the cantilever AC160TSA-R3 (300 kHz, 26 N/m, 7 nm tip radius). Images were post-processed with the in-built software features of IGOR 6.38801 (16.05.191, Asylum Research, Santa Barbara, CA, USA).\\
\indent\textbf{Raman Spectroscopy}\\
Raman polarization-dependent measurements were performed using a home-built spectroscopy setup \cite{shree2021guide}. The sample was placed inside a closed-cycle cryostat (Attocube systems, AttoDry 800). For above-band-gap excitation HeNe laser (E$_L$ = 1.96 eV, Thorlabs) and a green diode-pumped solid-state laser (E$_L$ = 2.33 eV, Roithner LaserTechnik) were used. The incident light was polarized using a Glan-Laser prism and then rotated with a Liquid Crystal Rotator placed in front of the objective (50x, NA = 0.7, CryoGlass Optics) on a low-temperature piezo-positioners (Attocube systems, ANPx101 and ANPz102) to position the sample with respect to the objective. The objective in the reflection configuration focused the laser beam on the sample surface at normal incidence. The spot size diameter was of the order of the wavelength used, confirmed experimentally for different wavelengths by scanning the reflection signal over a metal stripe. Spectral purity of the excitation light was ensured by using corresponding MaxLine filters (Semrock). Raman signal was collected by the same objective and sent via free space to the spectrometer (Teledyne Princeton Instruments IsoPlane300) coupled with a liquid nitrogen-cooled CCD (Teledyne Blaze 400HRX). The collected signal was dispersed with 1200 lines/mm diffraction grating. A spectral blocking of the Rayleigh were made with a corresponding long-pass filters (Verona Long-pass Raman Edge Filter, Semrock). The average power for all measurements was maintained in the range of 200-400 $\mu$W. All measurements were performed in a co-polarization configuration provided by placing a Glan-Laser polarizer before the spectrometer with the optical axis parallel to the Glan-Laser polarizer in the excitation path. \\
\indent\textbf{Differential reflectance contrast (DR/R) and photoluminescence excitation spectroscopy (PLE)}\\
DR/R and PLE spectroscopy were performed using the setup described in the previous section. As the excitation source we utilized a continuous white light fiber laser (Fianium FIU-15, NKT Photonics) coupled with a tunable high-contrast filter (LLTF Contrast VIS HP8, NKT Photonics) providing the spectral line width of $\sim$1 nm for PLE measurements or a broad line tunable filter (SuperK VARIA, NKT Photonics) for DR/R. For PLE measurement, the average incident power of 20 $\mu$W was maintained at each excitation energy from 1.46 to 2.76 eV with a gradient ND filter wheel and measured before the objective. The polarization of excitation light was set along the $a$ or $b$ crystallographic axis of a specific CrSBr flake. A 150 lines/mm diffraction grating was used.\\

\textbf{Data availability}\\
The data that support the findings of this study are available from the corresponding authors upon request.\\

 \textbf{Acknowledgements}\\
We thank Anton L\"ogl, Marc Anton Sachs, Gang Wang, and Alison Pfister for technical assistance. Z.S and K.M. were supported by ERC-CZ program (project LL2101) from the Ministry of Education, Youth and Sports (MEYS) and by the project Advanced Functional Nanorobots (reg. No. CZ.02.1.01/0.0/0.0/15.003/0000444 financed by the EFRR). CrSBr atomic structure was depicted using VESTA software \cite{momma2011vesta}.

\textbf{Author contributions} 
K.M. and Z.S. grew the bulk CrSBr crystals. D.I.M. and P.M. fabricated few-layer CrSBr samples for optical spectroscopy. D.I.M., S.S. and P.M. performed optical spectroscopy measurements. S.S installed the cryostat system. P.M., D.I.M, and R.v.K. performed and interpreted AFM measurements. D.I.M., L.K., and B.U. analyzed the optical spectra. M.M.G. contributed to the discussion and theoretical explanation of observed phenomena. All authors discussed the results. B.U. suggested the experiments and supervised the project. D.I.M., P.M., and B.U. wrote the manuscript with input from all the authors. \\

\textbf{Competing interests}: The authors declare no competing interests.\\

\indent \textit{Note added during submission }: While finalizing this manuscript, we became aware of very recent related work in \cite{zhai2025anisotropic}.


\begin{figure*}[t]
\includegraphics[width=1\linewidth]{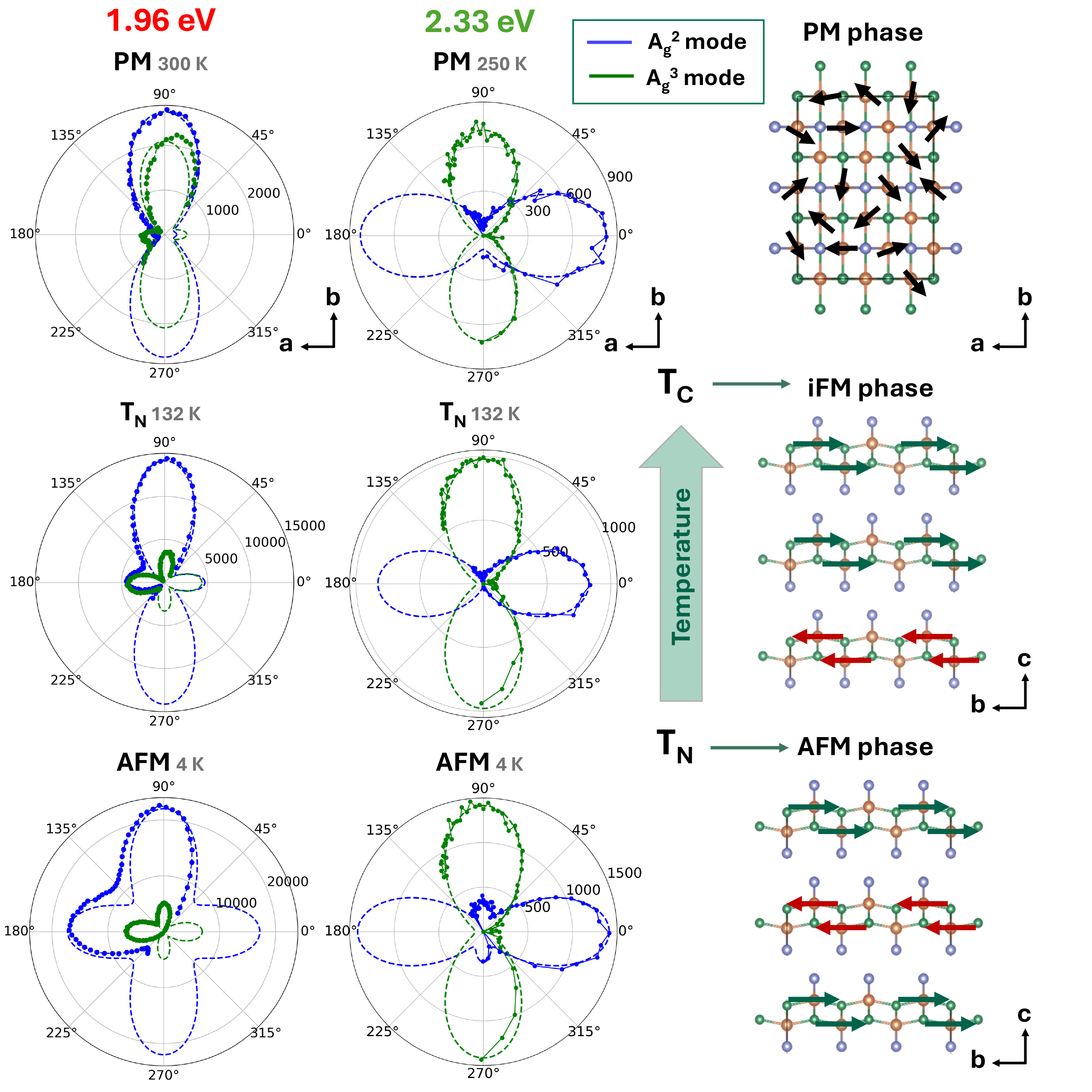}
\caption{\label{fig:fig1} \textbf{Temperature dependent evolution of polarization dependence for A$_g^2$ and A$_g^3$ Raman modes under different excitation energies in 5 nm thick CrSBr.} 
Polar plots display the Raman scattering intensity as a function of the polarization rotation angle for the A$_g^2$ mode (blue solid line) and the A$_g^3$ mode (green solid line) under excitations of 1.96 eV (left column) and 2.33 eV (center column). Measurements are shown for different temperatures (4 K, 132 K, 165 K, and either 250 K or 300 K), corresponding to distinct magnetic phases and characteristic transition temperatures. Solid lines with dots represent experimental spectra, dashed lines are fitting curves obtained using Eq.~\eqref{fit}. The polarization angle is measured relative to the $b$ axis (0$^o$). The radial axis indicates signal intensity, which is scaled independently for each plot, with the center corresponding to 0 counts. Right column: schematics of spin orientations in stacked CrSBr layers across different magnetic phases: (top) paramagnetic (PM) above 165 K, (middle) intermediate magnetic phase (iFM) between 132 K and 165 K (T$_C$), and (bottom) antiferromagnetic (AFM) below 132 K (T$_N$).}
\end{figure*}

\begin{figure*}
\includegraphics[width=1\linewidth]{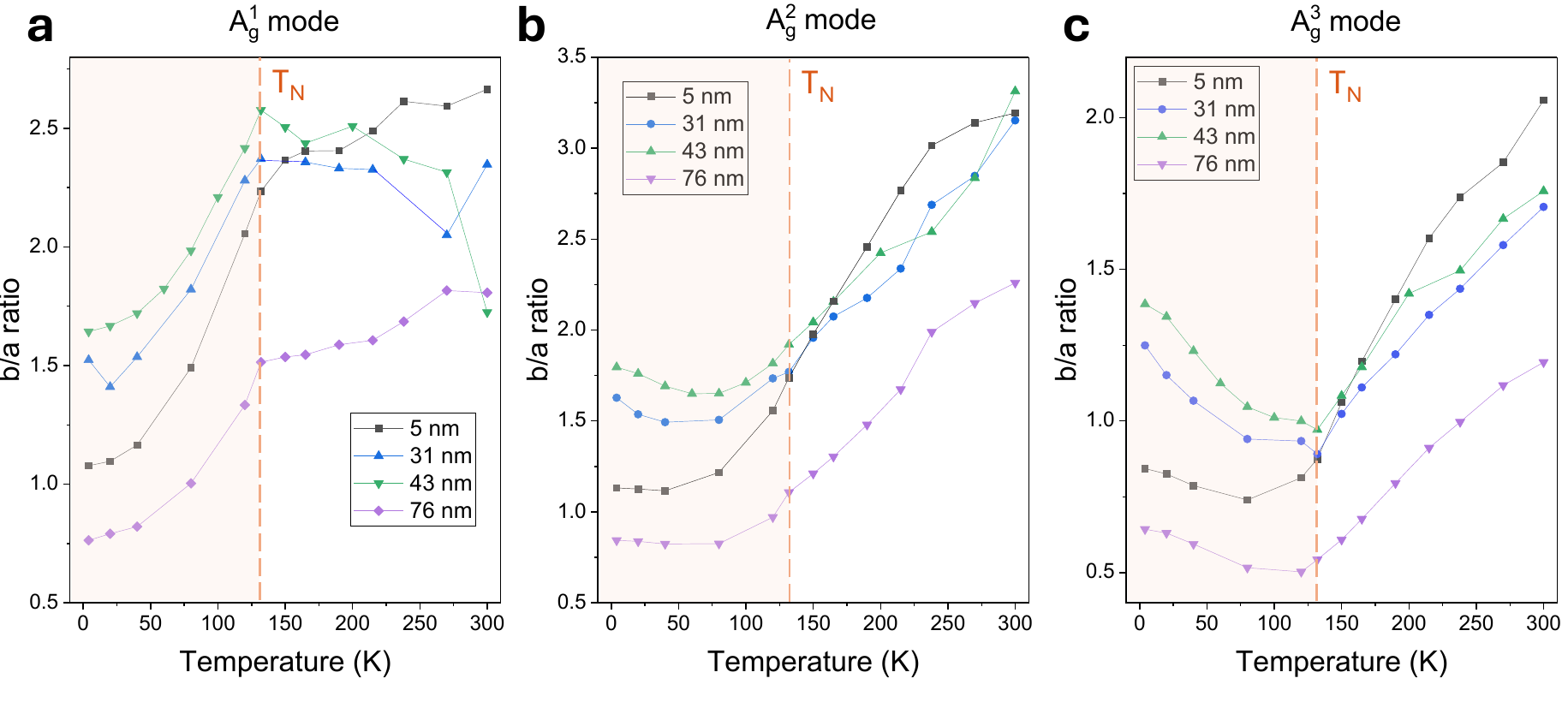}
\caption{\label{fig:fig2} \textbf{Temperature-dependent $b$/$a$ ratio of the A$_g^1$, A$_g^2$ and A$_g^3$ Raman modes under 1.96 eV excitation for different sample thicknesses.} Temperature-dependent $b/a$ ratio of the A$_g^1$, A$_g^2$, and A$_g^3$ Raman modes under 1.96 eV excitation for different sample thicknesses. Each Raman mode exhibits characteristic behavior near the Néel temperature across a broad thickness range. The orange-shaded region marks the antiferromagnetic phase, the orange dashed line indicates the Néel temperature (T$_N$ = 132 K). The $b$/$a$ ratios were extracted from fits of the polarization dependence using Eq.~\eqref{fit}.}
\end{figure*}

\begin{figure*}
\includegraphics[width=1\linewidth]{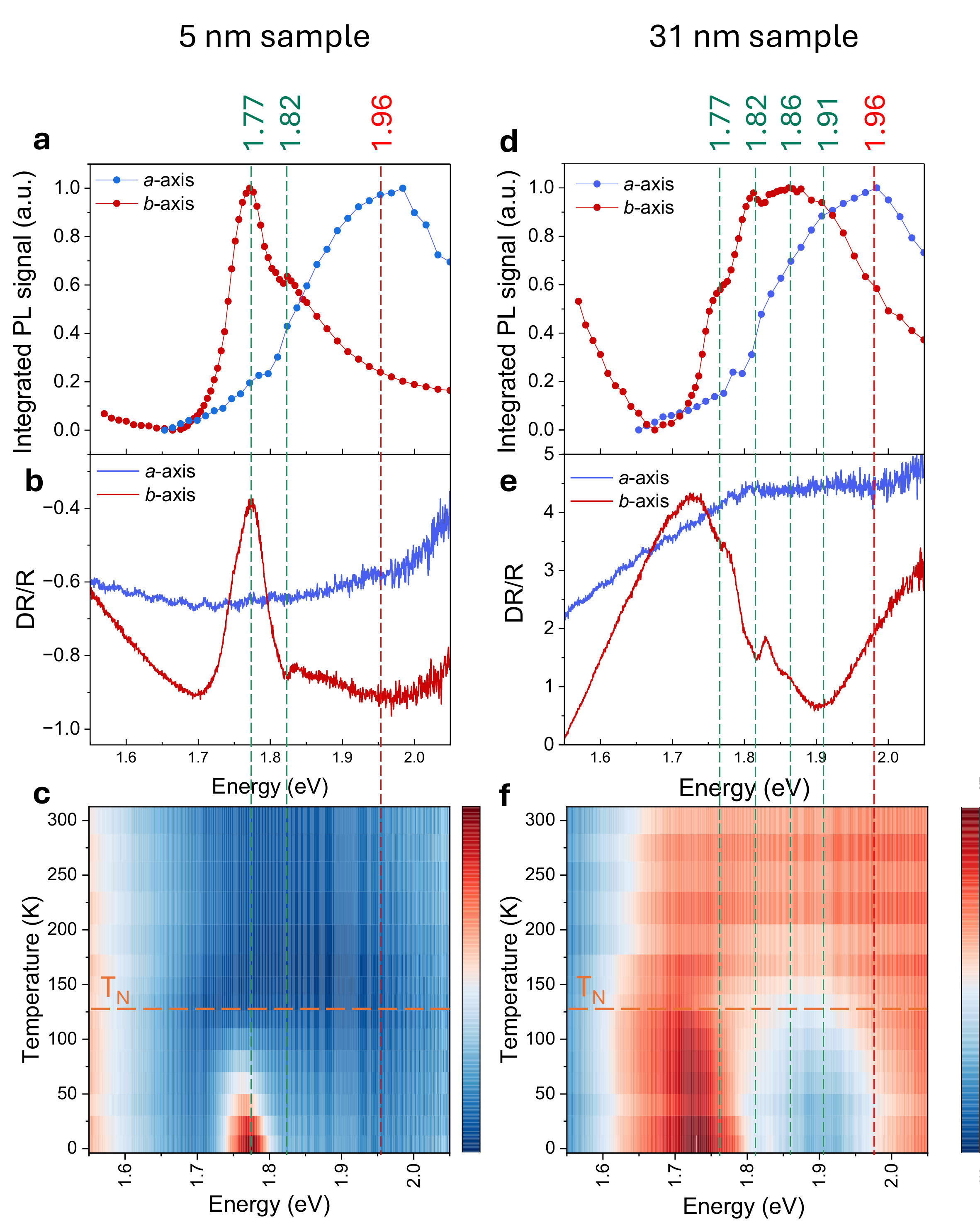}
\caption{\label{fig:fig3} \textbf{Differential reflectance contrast (DR/R) and photoluminescence excitation (PLE) spectroscopy.} (a, b) PLE and corresponding DR/R spectra for a 5 nm thick CrSBr flake. Blue curves represent excitation polarized along the $a$ axis, and red curves along the $b$ axis. The intensities along the vertical axes are normalized. Green and red dashed lines serve as guides to highlight matching spectral features. (d, e) Equivalent measurements for a 31-nm-thick flake. (c, f) False-color maps showing the temperature evolution of DR/R signal polarized along $b$-axis between 4 and 300 K for 5 nm (c) and 31 nm (f) thick flakes.}
\end{figure*}

\renewcommand{\thefigure}{S\arabic{figure}}
\setcounter{figure}{0}

\begin{figure*}
\includegraphics[width=1\linewidth]{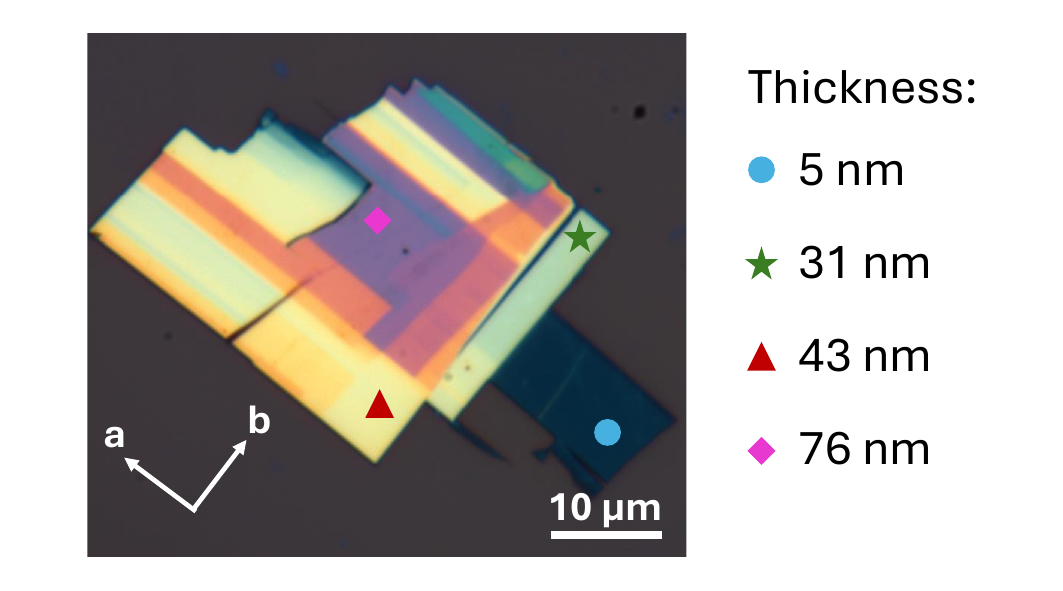}
\caption{\textbf{White light image of exfoliated flake}. Color marks indicate the regions of different thicknesses where measurements were taken. The thickness values were determined using atomic force microscopy (AFM).}
\label{fig:fig1} 
\end{figure*}
\newpage

\begin{figure*}
\includegraphics[width=1\linewidth]{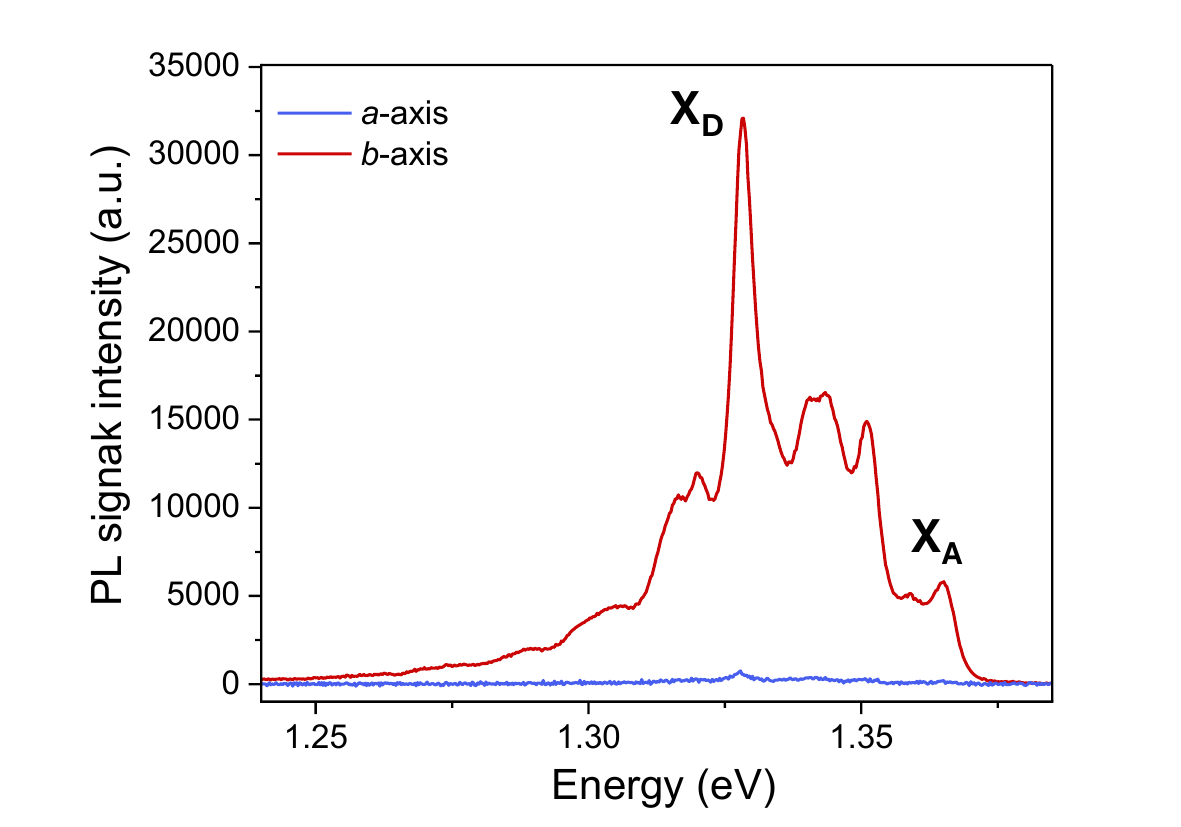}
\caption{\textbf{Photoluminescence spectra along $a$ and $b$ crystalographic axes.} Spectra were obtained from a 5 nm-thick sample under 1.96 eV excitation at 4 K. The spectrum along the $b$-axis exhibits the spectral features previously reported \cite{krelle2025magnetic}, namely the dark exciton X$_D$ at approximately 1.328 eV and the bright (A) exciton X$_A$ at around 1.365 eV.}
\label{fig:fig1} 
\end{figure*}
\bibliography{Ref} 

\newpage
\begin{figure*}
\includegraphics[width=1\linewidth]{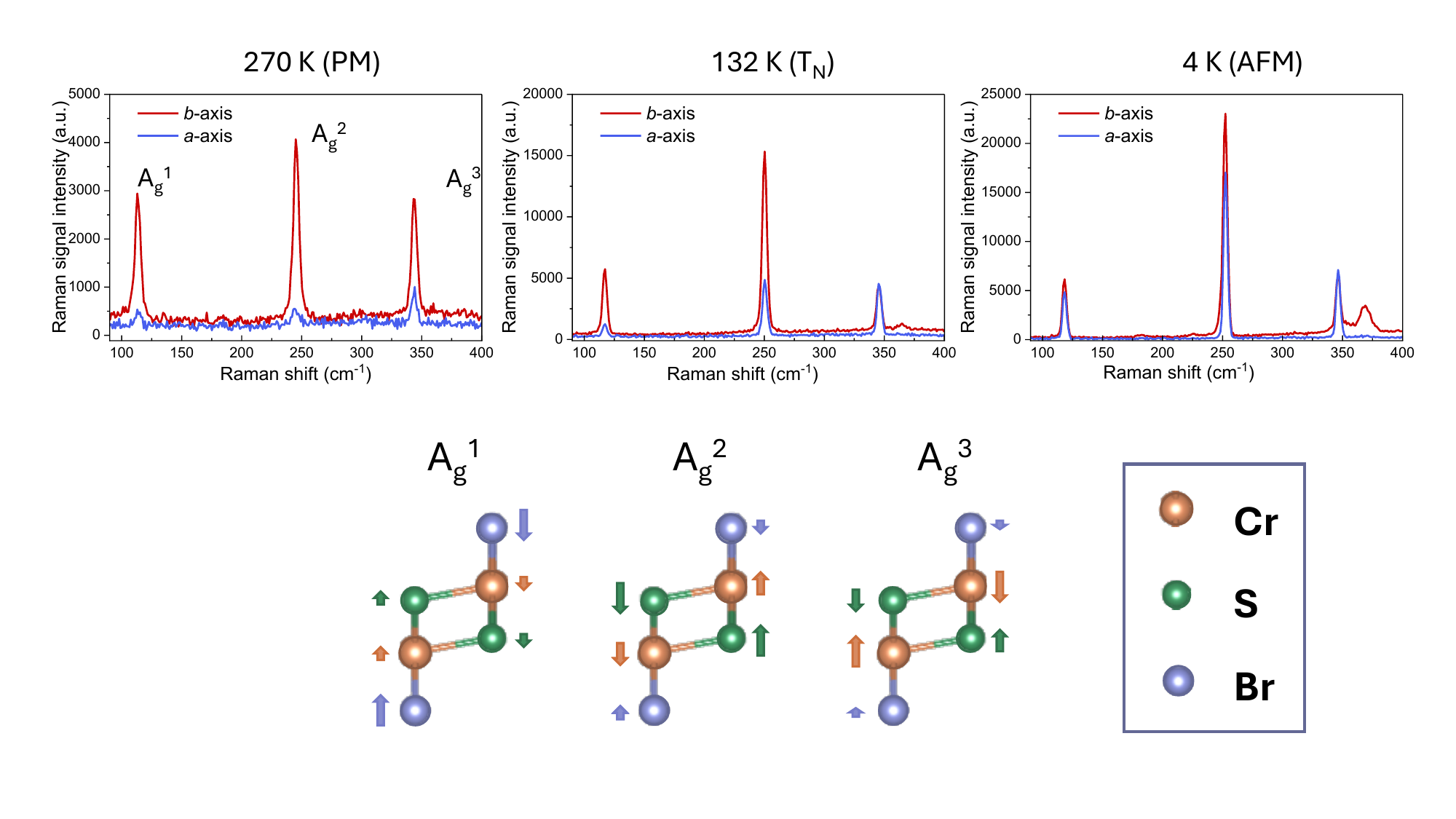}
\caption{\textbf{Raman spectra at different temperatures and corresponding Raman mode atomic motions.} Top panel: Raman spectra at various temperatures showing the behavior of the primary modes along the $a$- and $b$-crystallographic axes. Bottom panel: schematic of atomic displacements with relative amplitudes for different Raman modes. Orange spheres represent chromium, green represents sulfur, and blue represents bromine.}
\label{fig:fig1} 
\end{figure*}

\newpage
\begin{figure*}
\includegraphics[width=1\linewidth]{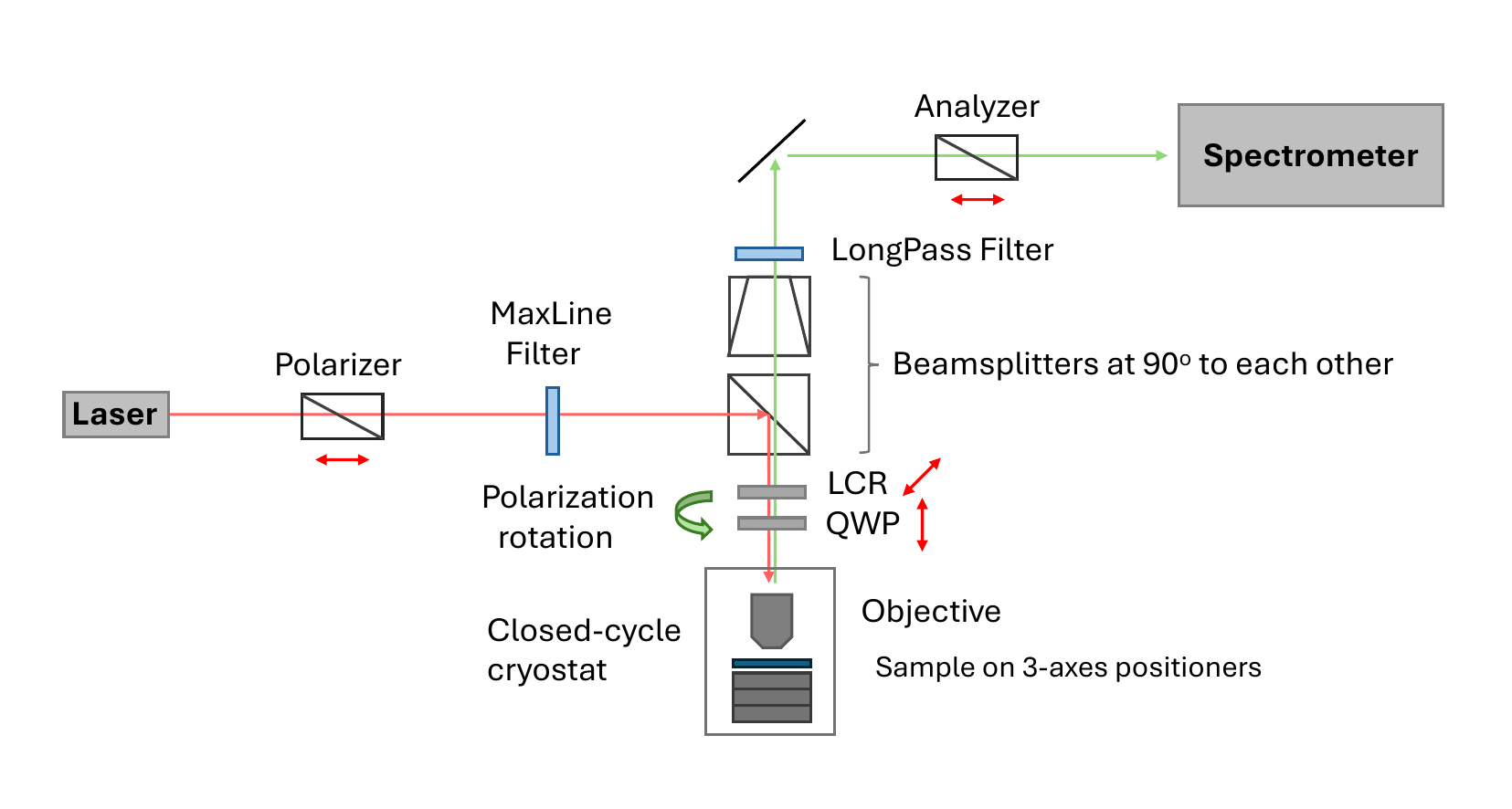}
\caption{\textbf{Experimental set-up.} In the image, LCR stands for liquid crystal retarder and QWP for quarter-wave plate. Red arrows indicate the orientation of the optical axis of each polarization element. For the polarizer and analyzer, the orientation is horizontal; the LCR optical axis is at 45° to the incident polarization, and the QWP is perpendicular to the incidence.}
\label{fig:fig1} 
\end{figure*}

\newpage
\begin{figure*}
\includegraphics[width=1\linewidth]{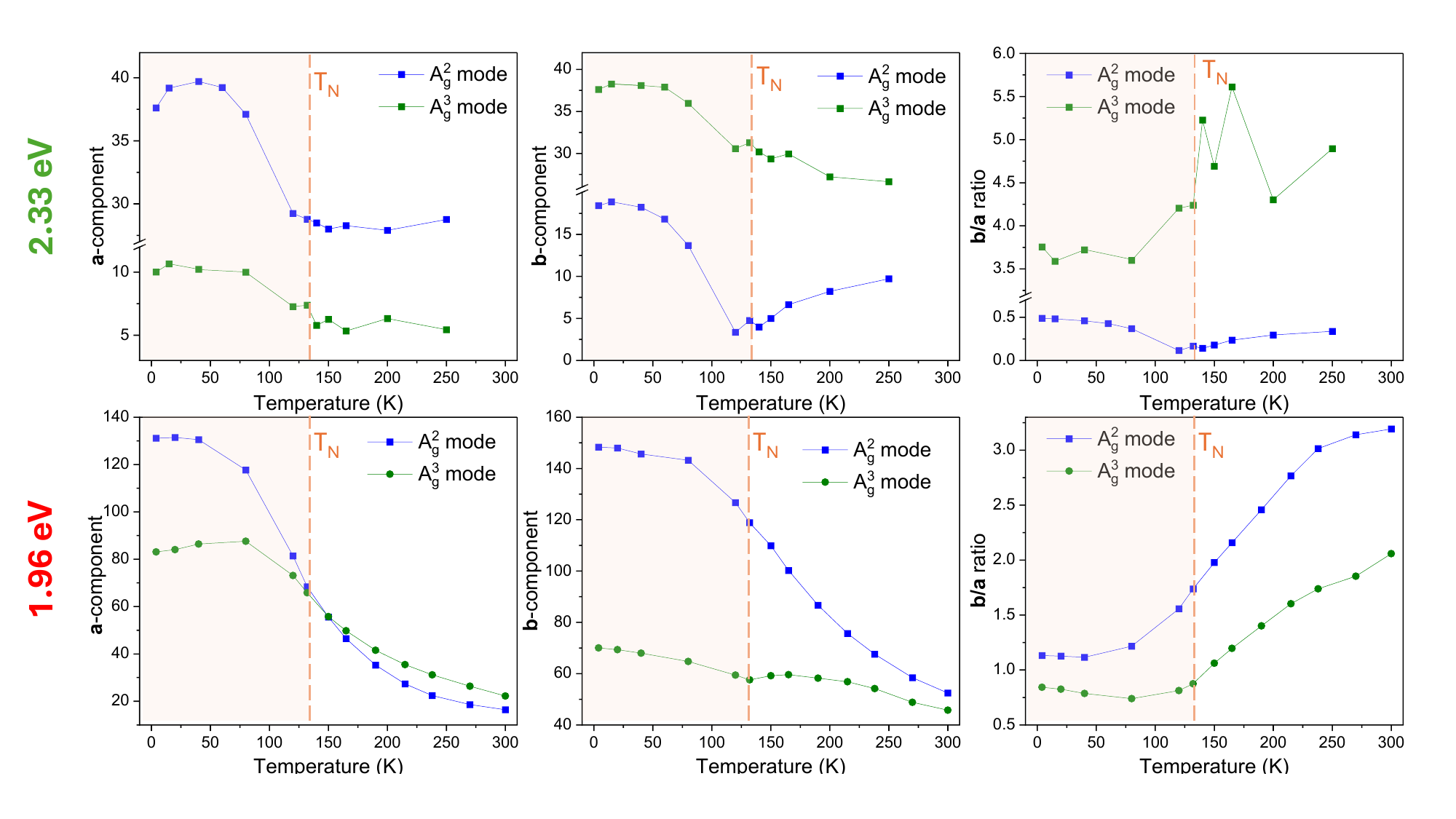}
\caption{\label{fig:fig3} \textbf{Temperature dependent Raman tensor elements (parameters, components) and b/a ratio of A$_g^2$ and A$_g^3$ Raman modes under different excitation energies.} (a-c) a-component, b-component, and b/a ratio under 2.33 eV excitation. (d-f) a-component, b-component, and b/a ratio under 1.96 eV excitation. Blue lines with dots correspond to A$_g^2$ mode, green lines with dots correspond to A$_g^3$ mode. The orange area indicates the temperature region of the antiferromagnetic phase, and the orange dashed line corresponds to the Neel temperature (T$_N$ = 132 K). Raman tensor elements were obtained from fitting the polar dependencies using Eq.(1)}
\end{figure*}

\newpage
\begin{figure*}
\includegraphics[width=1\linewidth]{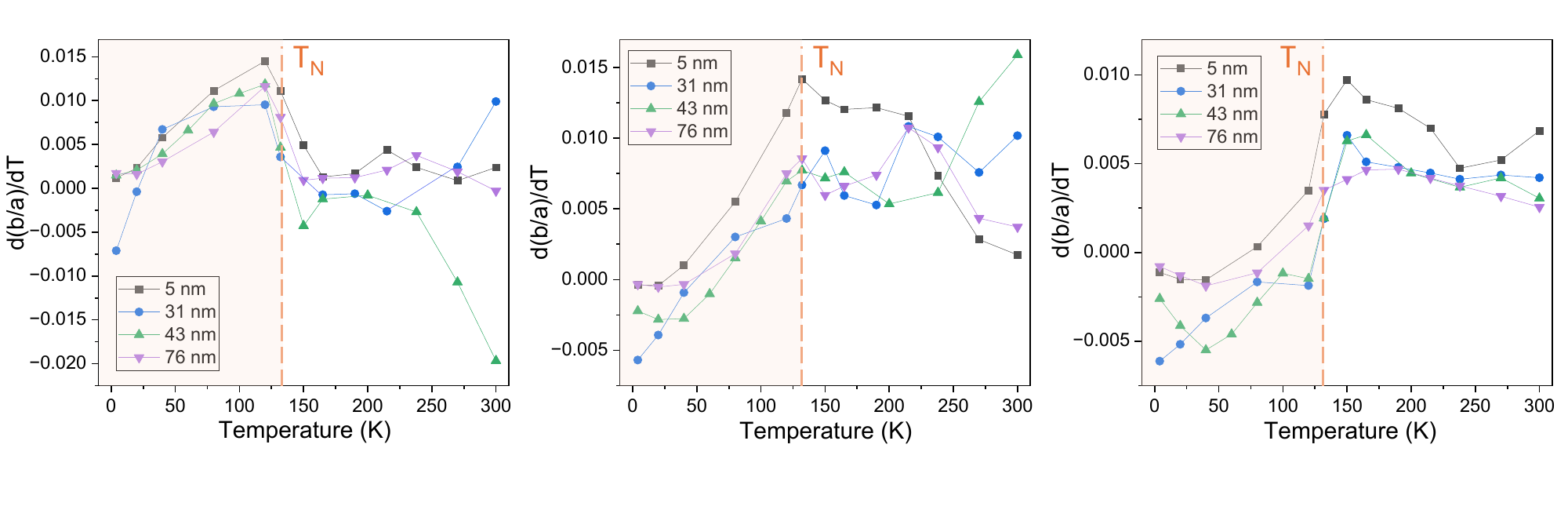}
\caption{\textbf{First derivatives of the b/a ratios for (a) A$_g^1$, (b) A$_g^2$ and (c) A$_g^3$ Raman modes.} The shaded orange area indicates the AFM magnetic phase.}
\label{fig:fig1} 
\end{figure*}

\newpage
\begin{figure*}
\includegraphics[width=1\linewidth]{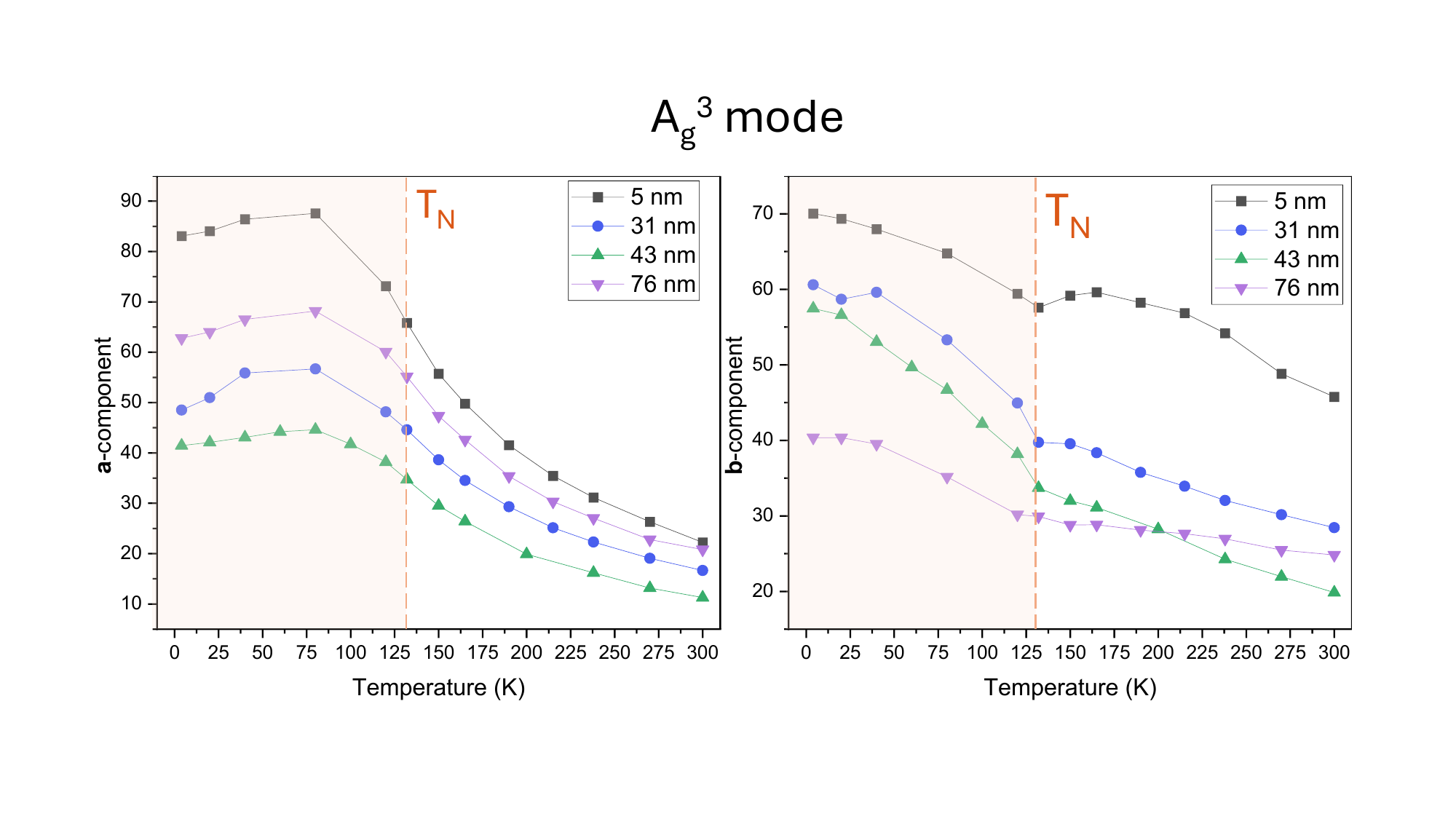}
\caption{\textbf{$a$- and $b$- components of A$_g^3$ mode for different sample thicknesses.} The shaded orange area indicates the AFM magnetic phase.}
\label{fig:fig1} 
\end{figure*}

\newpage
\begin{figure*}
\includegraphics[width=1\linewidth]{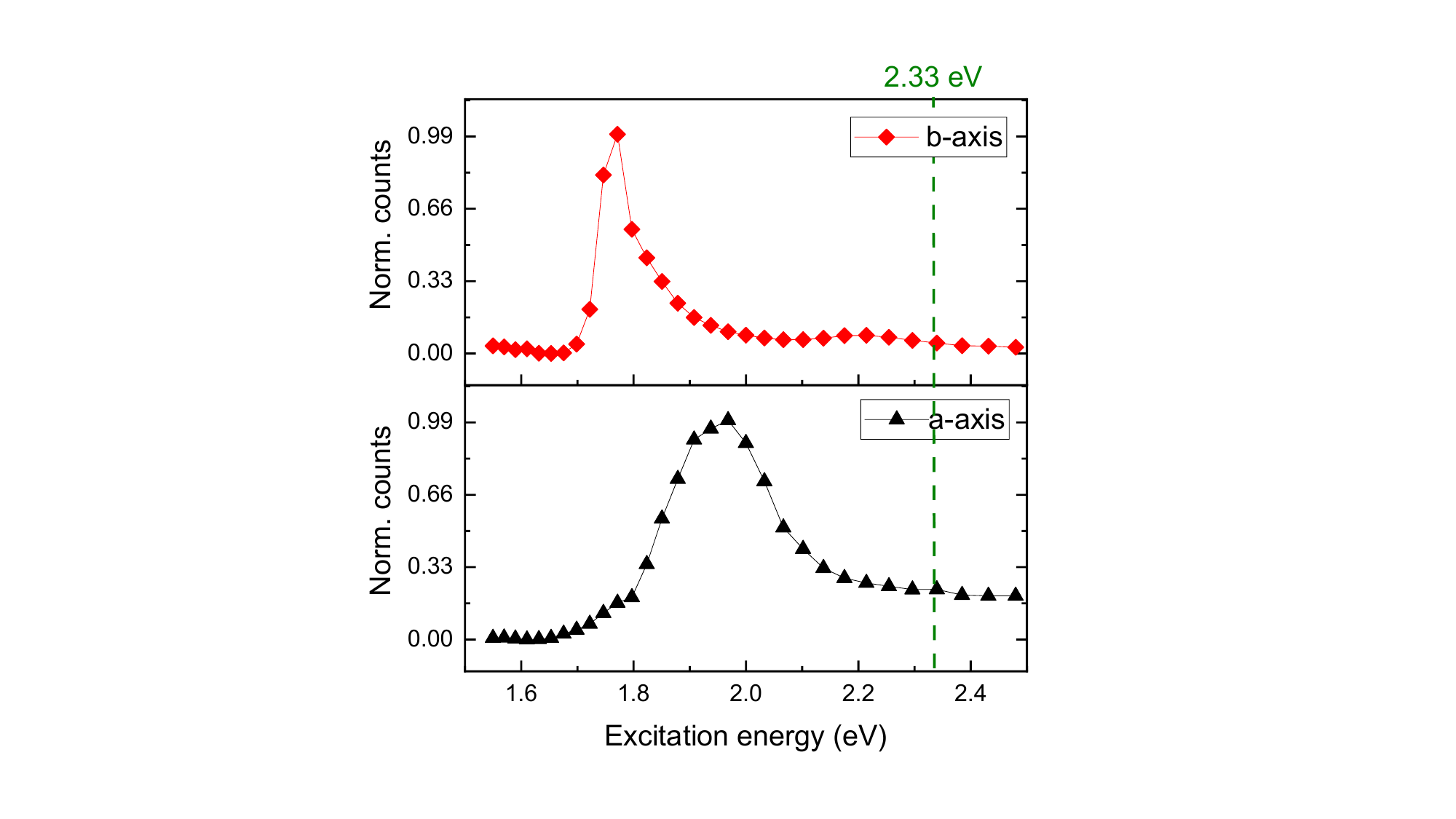}
\caption{\textbf{Extended Photoluminescence excitation (PLE) spectroscopy.} Measurements were performed on a 5 nm-thick sample at 4 K. Black curves correspond to excitation polarized along the $a$-axis, and red curves along the $b$-axis. Intensities on the vertical axes are normalized. Green dashed lines serve as guides, indicating 2.33 eV excitation.}
\label{fig:fig1} 
\end{figure*}

\newpage
\begin{figure*}
\includegraphics[width=1\linewidth]{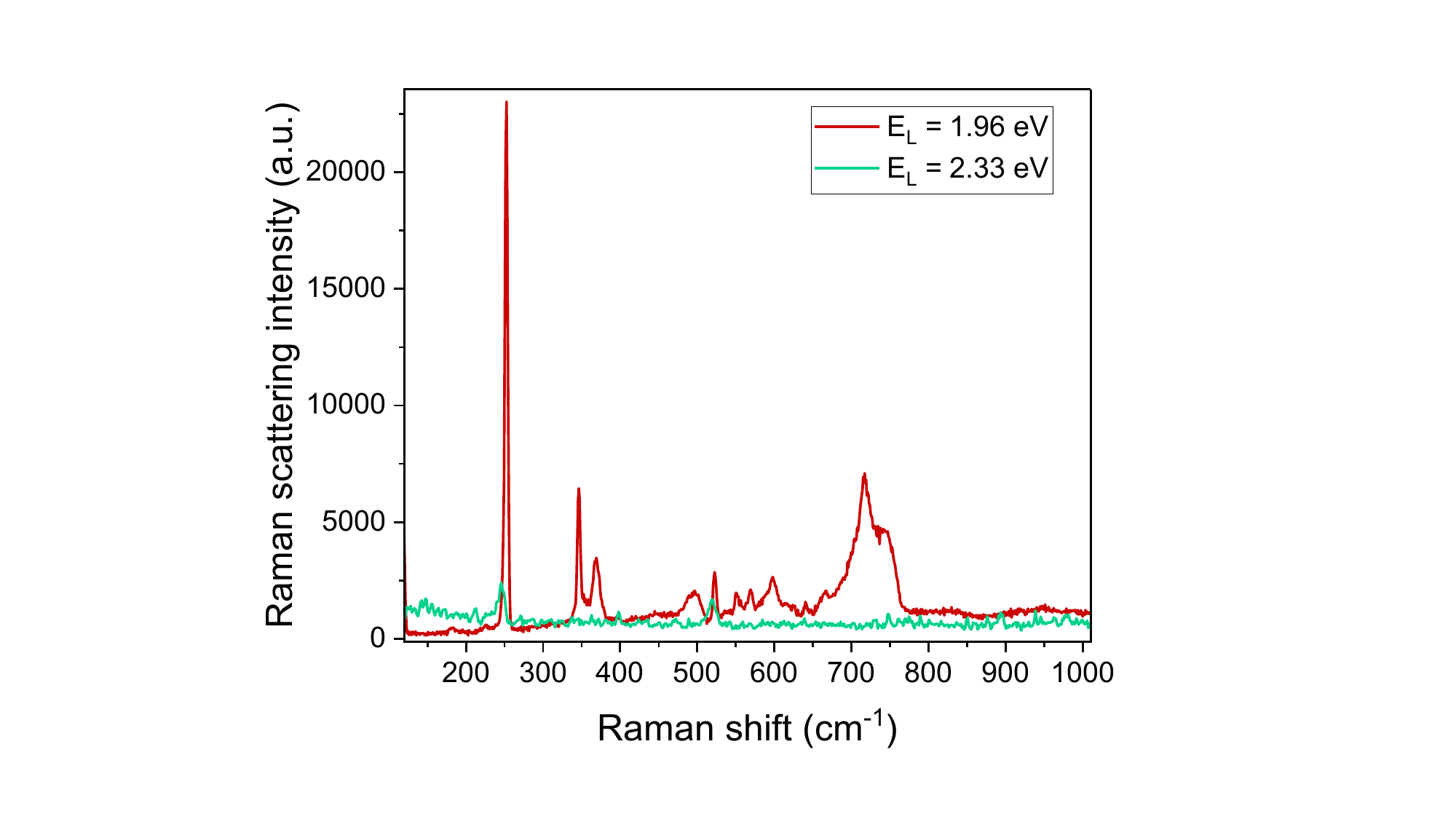}
\caption{\textbf{Expended Raman spectra for different excitation energies.} Spectra were obtained from a 5 nm-thick sample under 1.96 eV (red line) and 2.33 eV (green line) excitations at 4 K.}
\label{fig:fig1} 
\end{figure*}

\newpage
\begin{figure*}
\includegraphics[width=1\linewidth]{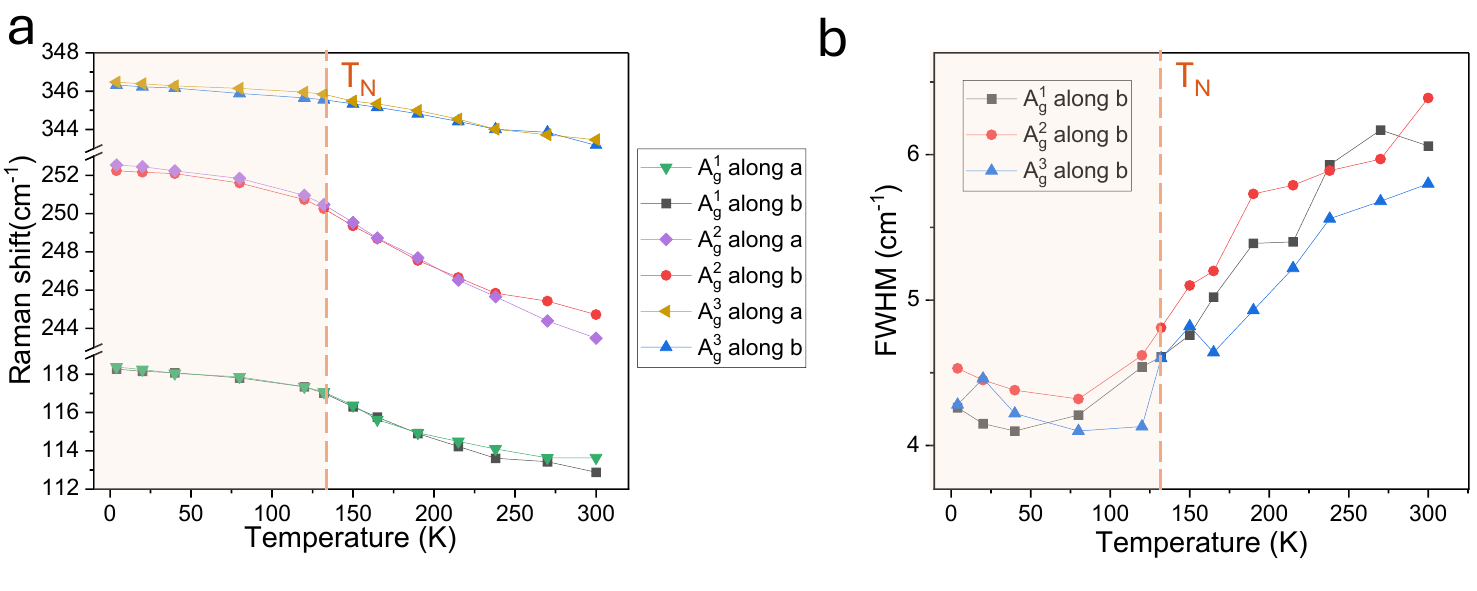}
\caption{\textbf{Temperature-dependent Raman mode parameters.} Temperature-dependent (a) Raman shift and (b) full width at half maximum (FWHM). Measurements were performed on a 5 nm-thick sample under 1.96 eV excitation. The shaded orange area indicates the AFM magnetic phase. All parameters were extracted by Gaussian fitting of the original data.} 
\label{fig:fig1} 
\end{figure*}

\end{document}